\documentclass[aps,prx,twocolumn,showpacs,superscriptaddress,floatfix,10pt,longbibliography,footinbib]{revtex4-2} 
\usepackage[utf8]{inputenc}
\usepackage[T1]{fontenc}
\usepackage{amsmath}
\usepackage{booktabs} 
\usepackage[caption = false]{subfig}
\usepackage[dvipsnames]{xcolor}
\usepackage{braket}
\usepackage{bm}
\usepackage{dsfont}
\usepackage{adjustbox}
\usepackage{graphicx}
\usepackage{enumitem}
\usepackage{bbm}
\usepackage{tikz}
\usepackage{physics}
\usepackage{microtype}

\usepackage{txfonts}

\usepackage[unicode=true,colorlinks=true,pdfpagemode=UseOutlines,pdfstartview=Fith]{hyperref}
\hypersetup{linkcolor=blue,citecolor=blue,urlcolor=blue}

\usepackage{slashed}
\usepackage{orcidlink}

\makeatletter

\usepackage{color}
\usepackage{xcolor}

\renewcommand{\Im}{\operatorname{Im}} 

\renewcommand{\Im}{\text{Im}}

\usepackage[normalem]{ulem}

\makeatother

\begin{document}


\title{Topological Triplons in the Pinwheel Valence Bond Solid on the Kagome Lattice\\}
\date{\today}

\author{L. Calonge-Martínez}
\affiliation{Technical University of Munich, TUM School of Natural Sciences, Physics Department, 85748 Garching, Germany}

\author{Peng Rao}
\affiliation{Technical University of Munich, TUM School of Natural Sciences, Physics Department, 85748 Garching, Germany}
\affiliation{Munich Center for Quantum Science and Technology (MCQST), Schellingstr. 4, 80799 M{\"u}nchen, Germany}

\author{F. Mila}
\affiliation{Institute of Physics, Ecole Polytechnique Federale de Lausanne (EPFL), CH-1015 Lausanne, Switzerland}

\author{J. Knolle}
\affiliation{Technical University of Munich, TUM School of Natural Sciences, Physics Department, 85748 Garching, Germany}
\affiliation{Munich Center for Quantum Science and Technology (MCQST), Schellingstr. 4, 80799 M{\"u}nchen, Germany}

\begin{abstract}
    We investigate the triplon excitations of the pinwheel valence-bond-solid phase on the deformed kagome lattice compound Rb$_2$Cu$_3$SnF$_{12}$. Using bond-operator mean-field theory, we compute the triplon band structure, dynamical structure factor, Berry curvatures and the associated thermal Hall response. We show that the presence of Dzyaloshinskii–Moriya interactions and an external magnetic field are key for endowing triplon bands with nontrivial Chern numbers. We find good qualitative agreement of the low-energy dynamical structure factor with neutron-scattering experiments. An applied magnetic field can isolate the lowest triplon Chern band leading to a tunable thermal Hall conductivity for accessible temperature and field regimes. Our results establish the deformed kagome pinwheel valence-bond solid as a promising platform for topological triplon physics and for observing the associated thermal Hall effect.
\end{abstract}
\maketitle

\section{\label{intro}Introduction}
Band topology has been successfully extended beyond electronic systems to a wide range of bosonic quasiparticles. Bosonic topology has been reported in a variety of systems: photons in topological photonic crystals, where analogues of quantum Hall physics were theoretically proposed and experimentally demonstrated~\cite{chiral_photon_1, chiral_photon_2,chiral_photon_3,chiral_photon_4}, phonons~\cite{zhang2010topological, zhang2011phonon,qin2012berry}, magnons in topological magnon insulators~\cite{chiral_magnon_1, chiral_magnon_2, katsura2010theory, chiral_magnon_4, mook2014magnon, chisnell2015topological, mcclarty2022topological} and, more recently, triplons in quantum magnets~\cite{romhanyi2015hall, hall_triplons_SS_2, hall_triplons_SS_3, d2024kitaev, esaki2505spin}, which are the focus of this work.
Triplons are spin-1 gapped bosonic excitations arising in valence-bond solid (VBS) phases. In a VBS, pairs of neighboring $S = 1/2$ spins form singlet states, resulting in a magnetically disordered ground state. The elementary excitation consists of breaking a singlet promoting it to a triplet state. Importantly, triplons can acquire nontrivial band topology. Moreover, in the context of magnetic materials, thermal Hall transport has emerged as a powerful experimental probe of topological bosonic excitations~\cite{zhang2024thermal} but its observation for quantum magnets with VBS phases and triplon excitations has remained elusive.

Theoretically, topological triplon bands were first reported on the Shastry–Sutherland lattice, where a combination of Dzyaloshinskii–Moriya interactions (DMI) and an external magnetic field opens topological gaps in the excitation spectrum~\cite{romhanyi2015hall, hall_triplons_SS_2, hall_triplons_SS_3}. Several works have predicted a finite thermal Hall effect of triplons in SrCu$_2$(BO$_3$)$_2$~\cite{romhanyi2015hall, malki2017magnetic, hall_triplons_SS_2, sun2021negative, bhowmick2021weyl}. In these systems, the magnetic field breaks time-reversal symmetry (TRS), while DMI break inversion symmetry, inducing chirality. The combination of both terms is responsible for complex hopping amplitudes of the triplons, which may then pick up Berry phases around closed paths, allowing for nonzero Chern numbers in the triplon bands. More recently, topological triplons have been investigated in other lattice geometries. On the Kitaev star lattice, it was shown that Heisenberg exchange together with a small magnetic field can give rise to nontrivial triplon topology~\cite{d2024kitaev}. Similar physics to that on the Shastry Sutherland lattice have been explored on the hexagonal and maple-leaf lattices~\cite{esaki2505spin}.

While spectroscpic signatures consistent with topological triplons have been reported for bulk probes, like inelastic neutron scattering on SrCu$_2$(BO$_3$)$_2$ ~\cite{hall_triplons_SS_2}, transport measurements (or surface sensitive probes) have not yielded evidence of topological triplon quasiparticles~\cite{suetsugu2022intrinsic, cairns2020thermal}. More specifically, Ref.~\cite{suetsugu2022intrinsic} investigated how the inter-triplon interactions influence the topological properties in SrCu$_2$(BO$_3$)$_2$ and demonstrated that the sign of thermal Hall coefficient is opposite compared to non-interacting predictions. Complementarily, Ref.~\cite{cairns2020thermal} reported no detectable thermal Hall signal of the expected magnitude within experimental resolution, suggesting that interaction effects and scattering processes significantly suppress coherent transverse transport in the investigated temperature regime. This discrepancy can be understood from the competing effects of a low quasiparticle density and interaction effects in SrCu$_2$(BO$_3$)$_2$. At low temperatures, the triplon density remains extremely small (of order $\sim 0.2\%$ around 5 K) because the gap of triplons is large compared to $k_B T$, resulting in very small thermal occupation of excitations to generate a sizable thermal Hall response. At higher temperatures, although the triplon population increases, interaction effects and thermal damping become significant, leading to a loss of quasiparticle coherence and a suppression of transport. As a result, there may be no accessible regime in which the simple non-interacting description applies quantitatively~\cite{suetsugu2022intrinsic, cairns2020thermal,habel2024breakdown,heinsdorf2025fate}.Therefore, it is highly desirable to find alternative triplon materials with a small, and ideally tunable, gap to the lowest excitation band.

In this work, we investigate the triplon excitations on a system based on the Rb$_2$Cu$_3$SnF$_{12}$ compound, which provides a realization of a pinwheel VBS on a kagome-based lattice. We show that the band structure in this system allows for a separation of the lowest triplon band from higher-energy excitations, where the gap is tunable via an applied magnetic field. This makes it a promising platform to explore topological properties of triplons and to assess the emergence of a thermal Hall response in a controlled setting. In particular, by analyzing the Berry curvature and associated transport coefficients, we establish conditions under which a finite thermal Hall conductivity can arise and is observable in realistic experimental conditions.

The compound Rb$_2$Cu$_3$SnF$_{12}$ was synthesized in Ref.~\cite{morita2008singlet}, where the ground state was identified as a disordered singlet phase with a finite spin gap and subsequent neutron scattering experiments and theoretical studies~\cite{matan2010pinwheel, mag_susceptibility, morita2008singlet, nikolic2003physics, yang2009valence} established that the system forms a 12-site pinwheel VBS ground state. In particular, Rb$_2$Cu$_3$SnF$_{12}$ exhibits a distorted kagome lattice. On the ideal kagome lattice, the triplet excitation spectrum is flat, consisting of localized and topologically trivial modes~\cite{yang2009valence, zeng1995quantum}. However, the lattice distortion lifts this degeneracy and stabilizes the 12-site pinwheel VBS phase when the deformed Heisenberg coupling is reduced to $\lesssim 0.97$ of the others~\cite{36-site, yang2009valence}, resulting in six triplon bands that are each threefold degenerate in the absence of a magnetic field. In addition, DMI, which are symmetry-allowed and experimentally expected in this compound~\cite{matan2010pinwheel}, further modify the excitation spectrum~\cite{hwang2012influence}.

Ref.~\cite{hwang2012influence} provided a comprehensive study of the triplon dispersion in the 12-site pinwheel VBS realized on the deformed kagome lattice. Using bond-operator mean-field theory, complemented by exact diagonalization and strong-coupling expansion, the authors computed the triplon excitation spectrum, the spin gap, the magnetic response and the effect of DMI on the magnetic excitations. However, despite the detailed understanding of the dispersion, the topological properties of the triplon excitations in this material have remained unexplored. This is especially intriguing given that the pinwheel VBS pattern is intrinsically chiral and that DMI act as an effective source of complex hopping amplitudes, both of which could be key ingredients for the emergence of nontrivial band topology. Experimentally, Ref.~\cite{matan2010pinwheel} provides a detailed characterization of the excitation spectrum using inelastic neutron scattering, observing no magnetic order down to 1.3 K and identifying low-energy gaps of $\Delta_1 = 2.35$ meV and $\Delta_2 = 7.3$ meV. These values are of the same order as in the Shastry–Sutherland compound SrCu$_2$(BO$_3$)$_2$ \cite{kageyama2000direct, mcclarty2022topological}. Importantly, Ref.~\cite{matan2010pinwheel} also reports dynamical structure factor measurements, providing a direct experimental observable against which our theoretical approach can be benchmarked.

Here, we build on the bond-operator analysis of Ref.~\cite{hwang2012influence}. First we review the model Hamiltonian and the bond-operator formalism in Sec.~\ref{Model}. Using the exchange couplings and DMI parameters extracted from the dimer expansion in Ref.~\cite{matan2010pinwheel}, we compute the triplon band structure in Sec.~\ref{Energy bands}. Going beyond previous studies, we evaluate dynamical and static structure factors in Sec.~\ref{DSF}, allowing for a direct comparison with neutron scattering experiments. This enables us to assess the validity of the bond-operator description for this system. In Sec.~\ref{sec:THE} we investigate the topological properties of the triplon bands by computing Berry curvatures and Chern numbers. We then make predictions for the associated thermal Hall response. In particular, at the level of the study of the triplon bands we address whether the bare intrinsic chirality of the pinwheel VBS pattern, as well as in combination with DMI and an external magnetic field, can give rise to nontrivial topology and measurable transverse thermal transport in realistic regimes. Our results provide experimentally testable signatures of topological triplon excitations in Rb$_2$Cu$_3$SnF$_{12}$.  

\begin{figure}[]
\centering
\includegraphics[width=1\linewidth]{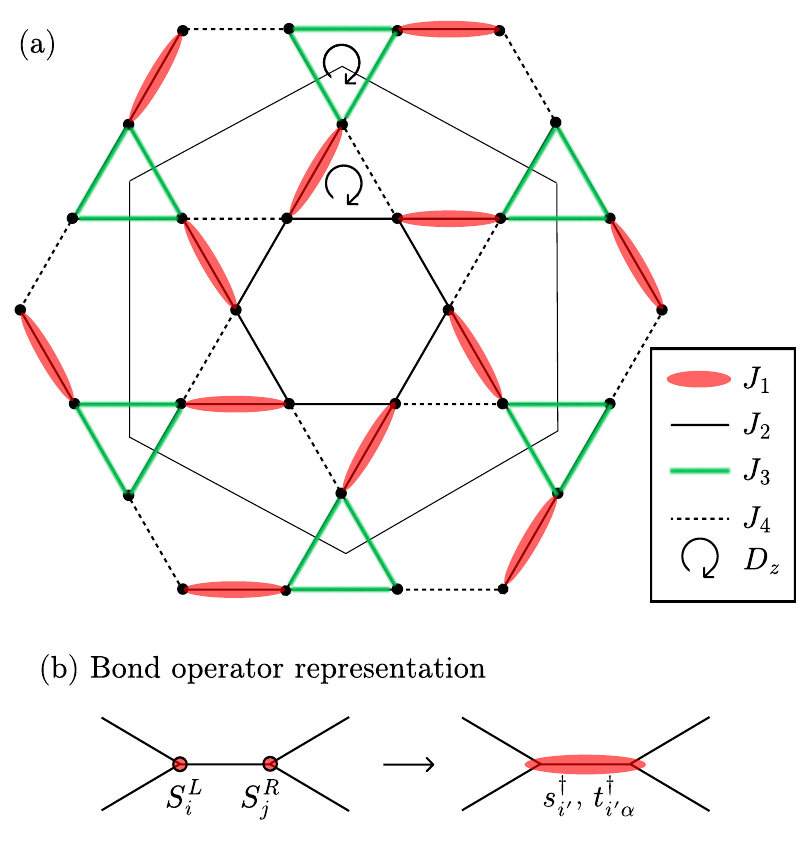}
\caption{\em (a) Deformed kagome lattice structure of Rb$_2$Cu$_3$SnF$_{12}$. The four inequivalent nearest-neighbor exchange couplings $J_1$–$J_4$ are indicated by distinct bond types, forming a 12-site pinwheel VBS unit cell. Red bonds denote singlet dimers, while arrows indicate the orientation of the out-of-plane DMI $D_{ij}^z$. (b) Schematic summary of the bond operator description.}
\label{deformed Kagome lattice} 
\end{figure}

\section{\label{Model} Lattice and Model}
\subsection{Lattice and Hamiltonian}
We start from an extended Heisenberg Hamiltonian including both DMI and an external magnetic field,
\begin{equation}
\mathcal{H} = \sum_{\langle ij \rangle} \Big[ J_{ij} \mathbf{S}_i \cdot \mathbf{S}_j + \mathbf{D}_{ij}^z \cdot (\mathbf{S}_i \times \mathbf{S}_j) \Big] + \mathbf{h} \cdot \sum_i \mathbf{S}_i,
\end{equation}
with $\mathbf{h} = g \mu_B \mathbf{B}$, $\mathbf{B}$ the applied magnetic field, $\mu_B$ the Bohr magneton, an estimated g-factor $g\approx2.4$~\cite{matan2010pinwheel}, and $\langle ij \rangle$ denotes nearest neighbour pairs.
The lattice is a deformed kagome lattice, where the four inequivalent nearest-neighbor exchange couplings $J_{ij}$ define distinct bond directions, as illustrated in Fig.~\ref{deformed Kagome lattice}(a). Specifically, the pairs of spins sitting on a $J_1$ coupled bond will be referred to as dimers. The unit cell can have two distinct dimer orientations, with the first configuration shown in Fig.~\ref{deformed Kagome lattice}(a) and the second one corresponding to exchanging the J$_1$ and J$_4$ bonds. These form a 12-site pinwheel pattern. The dimer pattern breaks translational and reflection symmetries, while preserving a $C_6$ rotational symmetry, and is intrinsically chiral, see Fig.~\ref{deformed Kagome lattice}(a), in the sense that the dimers have an orientation. We consider out-of-plane DM vectors, defined by their z-component $D_{ij}^z$, with a uniform sign convention corresponding to clockwise circulation around each triangle, as shown in Fig.~\ref{deformed Kagome lattice}(a). These interactions preserve the $C_6$ symmetry and encode a sense of chirality in the system, later playing a crucial role in shaping the triplon dispersion.

The exchange couplings and DMI strengths are taken from the dimer expansion analysis of Ref.~\cite{matan2010pinwheel}, which provides a quantitative description of Rb$_2$Cu$_3$SnF$_{12}$. Specifically, we use $J_1 = 18.6$ meV, $J_2 = 0.95 J_1$, $J_3 = 0.85 J_1$, and $J_4 = 0.55 J_1$, together with a dimensionless nearest neighbour DMI parameter $d_z = D_{ij}^z / J_{ij} = 0.18$, assuming that $D_{ij}^z$ scales proportionally with $J_{ij}$.
\subsection{Ground state, bond-operator formalism and mean-field treatment}
The ground state of the considered system is well described by a VBS ground state with a 12-site unit cell ~\cite{morita2008singlet, matan2010pinwheel, mag_susceptibility, nikolic2003physics, yang2009valence}. Specifically, the two spins sitting on the bonds with the strongest coupling $J_1$ and denoted in red in Fig.\ref{deformed Kagome lattice}(a), pair up to form a singlet state with gapped triplet excitations on top, referred to as triplons. A natural framework to describe this effective ground state and the excitations is the bond-operator formalism\cite{sachdev1990bond}. 

To analyze the excitations using the bond-operator formalism\cite{sachdev1990bond}, we follow closely the approach of Ref.~\cite{hwang2012influence} with minor modifications. In this framework, each singlet dimer is described in terms of bosonic operators creating singlet state, $|s\rangle$, and triplet states, $|t_{\alpha}\rangle$, out of vacuum, $|0\rangle$. In practice, spins sitting on a given singlet bond are labeled with $\mathbf{S}^L$ and $\mathbf{S}^R$, as in left and right spin forming the bond, and are treated as a single dimer site, as illustrated in Fig.\ref{deformed Kagome lattice}(b). This allows the ground state wavefunction to be approximated as a product of singlets on dimers, denoted by $\mathcal{D}$ :
\begin{equation}
    |\Psi_{GS} \rangle = \prod_{\mathcal{D}} |s\rangle_{\mathcal{D}}.
\label{singlet GS eq}
\end{equation}
Triplon excitations correspond to promoting a singlet into one of the three triplet states and can be treated as bosonic quasiparticles subject to a local hard-core constraint. The singlet–triplet states basis is given by
\begin{equation}
\begin{aligned}
\ket{s}   &= s^{\dagger} |0\rangle = \frac{1}{\sqrt{2}}(\ket{\uparrow \downarrow}-\ket{\downarrow \uparrow}), \\
\ket{t_x} &= t_x^{\dagger} |0\rangle =  \frac{i}{\sqrt{2}}(\ket{\uparrow \uparrow}-\ket{\downarrow \downarrow}), \\
\ket{t_y} &= t_y^{\dagger} |0\rangle =\frac{1}{\sqrt{2}}(\ket{\uparrow \uparrow} + \ket{\downarrow \downarrow}), \\
\ket{t_z} &= t_z^{\dagger} |0\rangle =\frac{-i}{\sqrt{2}}(\ket{\uparrow \downarrow}+\ket{\downarrow \uparrow}).
\end{aligned}
\label{bond_states}
\end{equation}
Note that this singlet–triplet basis differs slightly from that used in Ref.~\cite{hwang2012influence}, where triplets are organised by the spin projection quantum number. The two formulations are however equivalent. The present choice emphasizes the underlying SU(2) structure.

We now apply the bond-operator formalism to the present system. In the absence of DMI, the singlet–triplet basis defined in Eq.~\eqref{bond_states} diagonalizes the local Heisenberg dimer Hamiltonian. However, this is no longer the case when DMI are included and the local dimer Hamiltonian is given by 
\begin{equation}
    \mathcal{H}_{\text{Dimer}} = J_1 \mathbf{S}^L \cdot \mathbf{S}^R + \mathbf{D}_1 \cdot \mathbf{S}^L \times \mathbf{S}^R.
\end{equation}
For sufficiently weak DMI, the system remains in a quantum disordered phase, and the bond-operator description remains valid \cite{romhanyi2015hall}. However, DMI break SU(2) spin symmetry and require a modification of the local dimer basis given by Eq.~\eqref{bond_states}. To account for this, we perform a unitary rotation of the singlet and triplet states such that the local Hamiltonian on each dimer is diagonalized. For out-of-plane DM vectors, this results in a mixing between the singlet and the $t_z$ triplet component, 
\begin{equation}
\begin{aligned}
    |\tilde{s}\rangle &= \cos\theta |s\rangle - \sin\theta |t_z \rangle, \\
    |\tilde{t}_z\rangle &= \sin\theta |s\rangle + \cos\theta |t_z \rangle,
\end{aligned}
\label{bond_states rotated}
\end{equation}
with $\theta = \tfrac{1}{2}\arctan(d_z/J_1)$, while the transverse triplet modes remain unchanged, $|\tilde{t}_x\rangle = |t_x\rangle$ and $|\tilde{t}_y\rangle = |t_y\rangle$. The rotated basis $|\tilde{s}\rangle,|\tilde{t}_{\alpha}\rangle$ provides an effective description of the system in the presence of DMI and allows one to rewrite the spin operators and the Hamiltonian.

In this basis, the full Hamiltonian is rewritten as a sum of the following contributions:
\begin{equation}
    \mathcal{H} = \mathcal{H}_{\text{Dimer}}+ \mathcal{H}_{\mu} + \mathcal{H}_J + \mathcal{H}_D + \mathcal{H}_h .
\end{equation}
Specifically, the term $\mathcal{H}_{\mu}$ corresponds to the hard-core boson constraint, which is enforced via a Lagrange multiplier,
\begin{equation}
    \mathcal{H}_{\mu} = -\mu \sum_{i \in \mathcal{D}} (\tilde{s}^{\dagger}_i \tilde{s}_i + \tilde{t}_{i \alpha}^{\dagger} \tilde{t}_{i \alpha} -1),
\end{equation}
where $\mu$ is the Lagrange multiplier and is taken to be uniform across the unit cell, as it is invariant under 60 degrees rotation, implying that all the singlet pairs are equivalent.
The term $\mathcal{H}_h$ denotes the magnetic field contribution, which for a magnetic field in the z-direction is given by :
\begin{equation}
    \mathcal{H}_{h_z} = - i h_{z} \sum_{i \in D} \mathcal{E}_{z \alpha \beta} \tilde{t}_{i \alpha}^{\dagger} \tilde{t}_{i\beta}.
\label{mag field}
\end{equation}
The contribution from the inter-dimer Heisenberg couplings is of the form :
\begin{equation}
\begin{split}
& \mathcal{H}_{J} = \sum_{\langle ij \rangle '} \sum_{\alpha= x,y,z}\Bigg[C_{ij}^{\alpha} ( \tilde{s}_i\tilde{s}_j^{\dagger}\tilde{t}^{\dagger}_{i \alpha}\tilde{t}_{j \alpha }-\tilde{s}_i\tilde{s}_j\tilde{t}^{\dagger}_{i \alpha }\tilde{t}^{\dagger}_{j \alpha } + h.c.) \\ &
+ \frac{1}{2}\sum_{\beta \neq \alpha} B_{ij}^{\alpha}(\tilde{t}^{\dagger}_{i \alpha }\tilde{t}^{\dagger}_{j \alpha}\tilde{t}_{i \beta}\tilde{t}_{j \beta } - \tilde{t}^{\dagger}_{i \alpha }\tilde{t}_{j \alpha }\tilde{t}_{i \beta }\tilde{t}^{\dagger}_{j \beta } + h.c.)  \Bigg]
\end{split}
\end{equation}
where $C_{ij}^{\alpha}$ and $B_{ij}^{\alpha}$ are constants that depend on $\theta$. Similarly, $\mathcal{H}_D$, which corresponds to the inter-dimer contribution of the DMI term, has an analogous form. For both $\mathcal{H}_J$ and $\mathcal{H}_D$ there are additional cubic terms such as $\mathcal{E}_{ \alpha \beta\gamma}s^{\dagger}t_\alpha t_\beta^\dagger t_\gamma$. For the bare Heisenberg Hamiltonian these terms vanish after mean field decoupling since $P_{ij},Q_{ij} \propto \delta_{ij}$, see Eq.~\eqref{MF} below for the definition of the mean fields. In the presence of DMI these terms do not generally vanish but hybridize the singlet and triplet sectors. For small DMI however, we do not expect these terms to qualitatively affect the triplon spectrum and will neglect them in this paper. The full expressions of $\mathcal{H}_{\text{Dimer}}$, $\mathcal{H}_J$ and $\mathcal{H}_D$ are given in Appendix~\ref{app:bondop}.

Quartic interaction terms are treated within a self-consistent mean-field approximation. Within the mean-field theory, the hard-core constraint is enforced on average via the Lagrange multiplier and the singlet operator is replaced by its expectation value $\bar{s}$, assumed uniform due to the $C_6$ symmetry of the unit cell, with $\bar{s}=1$ corresponding to a fully condensed singlet background. The quartic terms are decoupled by introducing the following mean fields:
\begin{equation}
\begin{aligned}
    P_{ij}^{\alpha \beta} (\mathbf{R}) &= \langle t^{\dagger}_{i \alpha }(\mathbf{r})t_{j \beta }(\mathbf{r} + \mathbf{R}) \rangle,
    \\ Q_{ij}^{\alpha \beta}(\mathbf{R}) &= \langle t_{i \alpha }(\mathbf{r}) t_{j \beta }(\mathbf{r} + \mathbf{R}) \rangle,
\end{aligned}
\label{MF}
\end{equation}
where $\mathbf{R}$ denotes the relative displacement between unit cells and the expectation value is taken over the singlet ground state given in Eq.~\eqref{singlet GS eq}. In practice, the resulting mean fields are found to be small, with typical values around $0.02$. Consequently, mean field contributions from triplon-triplon interactions are subleading compared to the dominant contributions proportional to $\tilde{s}^2$, which is of order unity. This procedure yields an effective quadratic Hamiltonian, whose parameters, including $\mu$ and $\bar{s}$, are determined self-consistently from the coupled saddle-point equations.

The resulting quadratic Hamiltonian is Fourier transformed to momentum space, yielding a bosonic problem with an $18 \times 18$ matrix structure. Due to the presence of anomalous terms, the Hamiltonian is diagonalized using a bosonic Bogoliubov–de Gennes (BdG) method. The resulting paraunitary Bogoliubov transformation then provides both the triplon dispersion relations and the corresponding eigenvectors. These quantities serve as the basis for computing dynamical structure factors as well as topological properties such as Berry curvature and Chern numbers.

Further technical details of the formalism, including the full Hamiltonian and self-consistency equations, are provided in Appendix~\ref{sec:triplonMF}.
\section{\label{Energy bands} Triplon dispersion bands}
\begin{figure*}[t]
  \centering
  \includegraphics[width=1\textwidth]{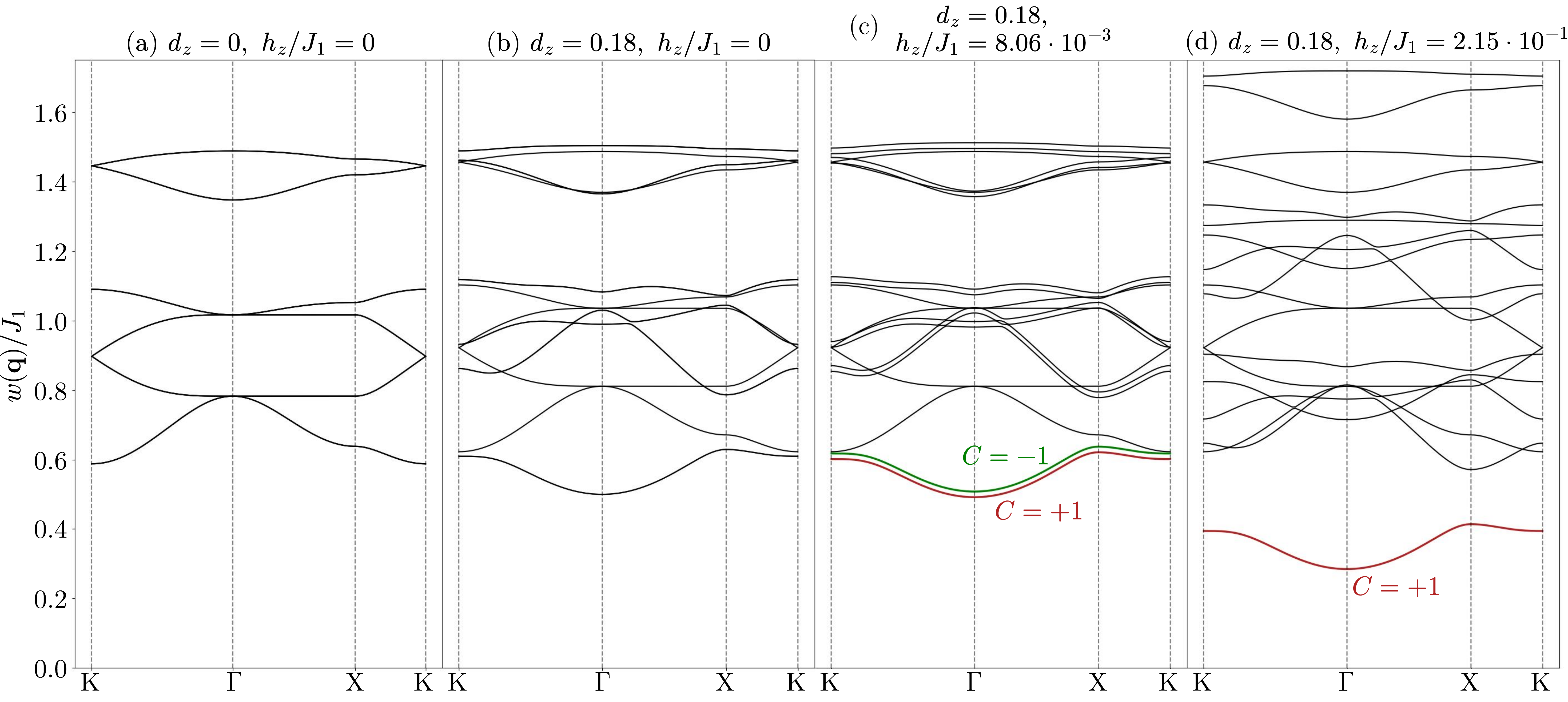}
  \caption{\em Triplon energy bands for the nearest-neighbor pinwheel valence-bond solid on the kagome lattice. (a) Spectrum in the absence of DMI and magnetic field, showing threefold degenerate triplet modes. (b) Inclusion of DMI with $d_z = 0.18$. (c) Addition of a small magnetic field further splits the transverse ($x$ and $y$) triplon branches. (d) Larger magnetic field leading to a clear isolation of the lowest-energy band.}
  \label{energy bands}
\end{figure*}
The full excitation spectrum consists of 18 modes, as there are six dimers per unit cell and three triplet channels. In the absence of DMI and magnetic field, the three triplet channels remain degenerate, reducing the spectrum to six bands that are each threefold degenerate, as shown in Fig.~\ref{energy bands}(a).

The band structure exhibits degeneracies at special momenta points of the Brillouin zone (BZ), as shown in Fig.~\ref{energy bands}(a). These degeneracies can be understood from symmetry considerations. Indeed, a systematic way to classify the bands is through group-theoretical arguments \cite{yang2009valence}. The unit cell retains a $C_6$ rotational symmetry while breaking reflection symmetry with respect to the $\Gamma$ point. As $C_6$ is an abelian group, its irreducible representations (irreps) are one-dimensional and labeled by $\rho_m(C_6)=e^{i\frac{2 \pi m}{6}}\text{, } m=0,1,2,3,4,5$. While this would suggest six nondegenerate bands, the bare Heisenberg Hamiltonian is real-valued, such that complex-conjugate irreps pair up, leading to twofold degeneracies. In particular, $\rho_1$ and $\rho_5$, as well as $\rho_2$ and $\rho_4$, form conjugate pairs, producing degeneracies at the $\Gamma$ point. Similarly, the $C_3$ rotational symmetry with respect to the $K$ point, leads to irreps $\rho_0=1$ and $\rho_{1,2}=e^{\pm i2\pi/3}$, where the latter are complex-conjugates, which will yield a two-fold degeneracy at the $K$ point. These degeneracies are not symmetry-protected but arise from the reality condition of the Hamiltonian. As a consequence, they can be lifted by perturbations, such as DMI, that render the Hamiltonian complex without needing to break the underlying lattice symmetry.

A natural question is whether the intrinsic chirality of the pinwheel VBS pattern can by itself induce topological triplon bands, without DMI, only adding a magnetic field to break TRS. Applying a magnetic field lifts the triplet channel degeneracy via a Zeeman splitting, linearly separating the three triplon branches with an energy shift. However, from Eq.~\eqref{mag field} we see it does not couple different dimers and importantly, this effect acts uniformly across the spectrum, regardless of band index. Indeed, under a finite magnetic field, which can always be taken to be along the $z$-axis due to spin rotation symmetry, the term \eqref{mag field} in the Hamiltonian can be diagonalised by triplon components $t_z, t_\pm \sim t_x \pm i t_y$ into $\pm h \sum_i t_{\pm,i}^\dagger t_{\pm,i}$. This simply removes the three-fold degeneracy of the triplets and introduces uniform energy shifts $0, \pm h$ to each band. As a result, it cannot lift the accidental degeneracies at both the K and the $\Gamma$ points. From a topological perspective, it means that although the magnetic field breaks TRS, it does not open topological gaps in the spectrum, or in other words, it does not induce a topological phase transition. Therefore, while such splitting is necessary to isolate triplon bands, it is insufficient to induce nontrivial band topology.

The effect of DMI on the triplon bands has been analyzed in Ref.\cite{hwang2012influence}. Importantly, DMI leads to gap closings and reopenings in the spectrum, indicating potential topological phase transitions. For the value $d_z = 0.18$, the lowest band softens at the $\Gamma$ point and approaches a near-degeneracy at the $K$ point, as shown in Fig.~\ref{energy bands}(b). This is in agreement with neutron scattering results \cite{matan2010pinwheel}. On the other hand, the spin gap is found to be approximately $0.5 J_1$. This result is, as noted in Ref.\cite{hwang2012influence}, significantly larger than the experimental value $0.126J_1$ reported by \cite{matan2010pinwheel}. This discrepancy can be attributed to the fact that triplon mean field theory is only qualitative. We argue that the mean field theory, will tend to overestimate the spin gap in dimer systems, as the singlet background can be over-stabilized and quantum fluctuations, triplon-triplon processes and virtual processes that delocalize excitations, are only partially captured and could lower the energy. Furthermore, the out-of-plane DMI partially lift the threefold degeneracy of the triplet channels, splitting the bands into two distinct branches. In particular, the degeneracies at the $\Gamma$ and $K$ points are lifted for the transverse ($x,y$) channels.

Finally, we consider the effect of a magnetic field applied along the $z$ direction. This further splits the transverse $x$ and $y$ channel bands, allowing individual bands to be energetically isolated, as illustrated in Fig.~\ref{energy bands}(c). The lowest band can be further lowered and isolated, see Fig.~\ref{energy bands}(d), where the field shown corresponds to approximately $30$~T. The critical field at which the lowest triplon mode condenses, signaling the onset of magnetic order, can be directly estimated from the spin gap and is found to correspond to about $67$~T. This result is again larger than the experimental value $21$~T reported by \cite{matan2010pinwheel}. We argue this can in part be attributed to the spin gap being overestimated, and this result should therefore be interpreted qualitatively. Nevertheless, the order of magnitude is accurate and the field-induced splitting of the transverse modes is crucial for defining well-separated bands, enabling the analysis of Berry curvature and associated topological properties.

\begin{figure*}[t] 
\centering
\includegraphics[scale=0.37]{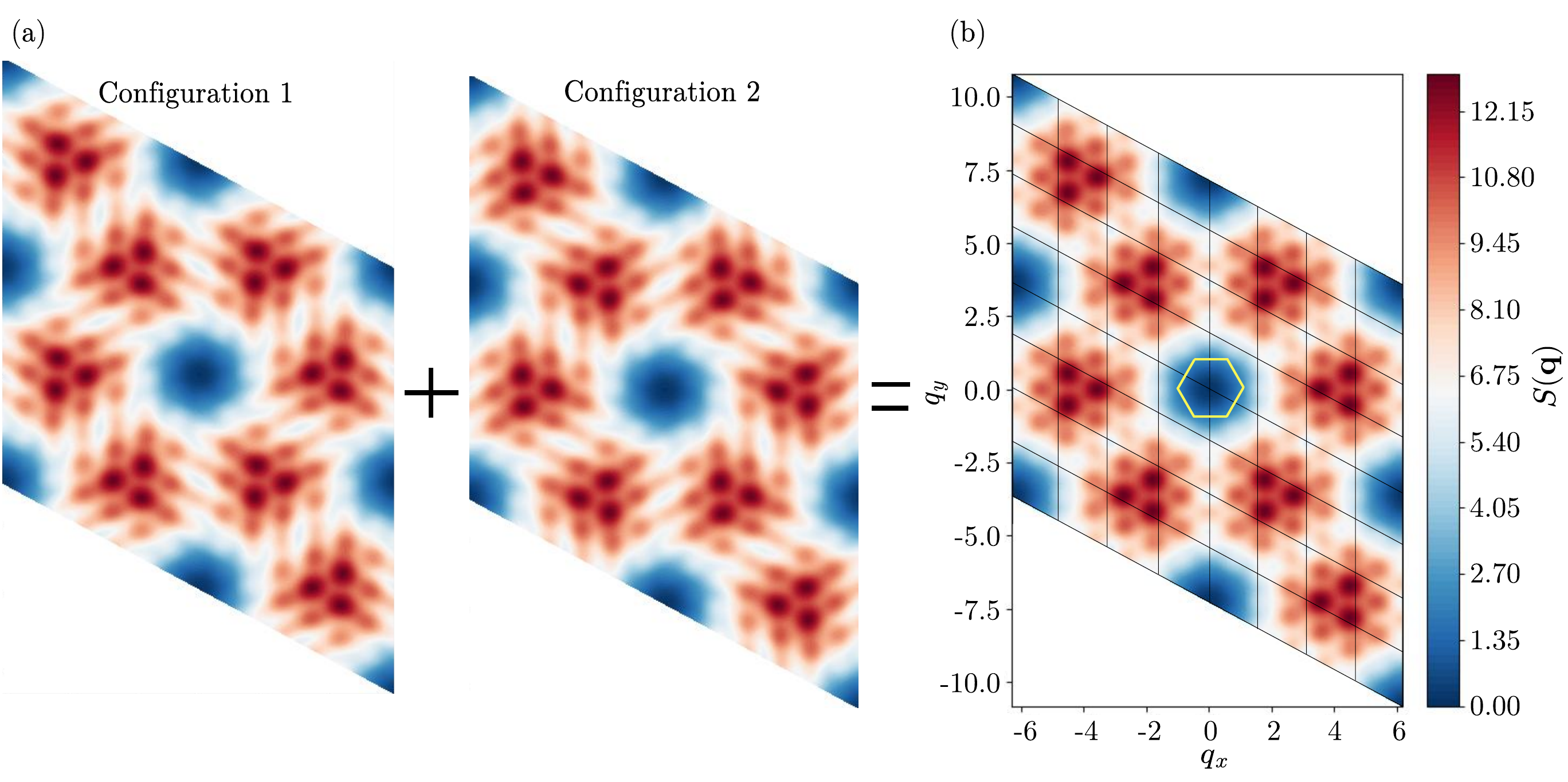}
  \caption{\em Static structure factor $S(\mathbf{q}) = S^{xx}(\mathbf{q})+S^{yy}(\mathbf{q})+S^{zz}(\mathbf{q})$ for $d_z = 0.18$ and zero magnetic field, obtained by assembling 64 reduced Brillouin zones. (a) The two panels correspond to the two chiral configurations. (b) $S(\mathbf{q})$ averaged over both configurations. The background grid indicates the tiling of the the 64 shifted reduced Brillouin zones used in the construction, while the hexagon highlighted in yellow marks the first Brillouin zone}
\label{SSF}
\end{figure*}
\section{\label{DSF}Structure Factor}
Within the bond-operator formalism, the approximate ground state and its associated correlation functions allow for an analytical evaluation of both dynamical and static structure factors. This enables a direct comparison between the theoretical triplon description and the INS experiments reported in Ref.~\cite{matan2010pinwheel} for Rb$_2$Cu$_3$SnF$_{12}$. In addition, the static structure factor provides further insight into the underlying effective ground-state correlations within the bond-operator framework and the lattice.
\subsection{Static Structure Factor}
The static structure factor is defined as
\begin{equation}
    S^{\alpha \beta}(\textbf{q}) = \frac{1}{N_s} \sum_{m,n} e^{i \textbf{q}(\textbf{r}_n - \textbf{r}_m)} \langle {S}^{\alpha}_m {S}^{\beta}_n \rangle,
\label{eq struct fact}
\end{equation}
where $\mathbf{r}_n$ denotes the position of the nth spin, $N_s$ is the number of site and $\alpha,\beta =x,y,z$. Within the bond-operator formalism, the spin operators in Eq.~(\ref{eq struct fact}) can be expressed in terms of bosonic triplon operators. We label the six dimers within the unit cell by indices $i',j'$, and denote by $a,b \in {L,R}$ the two spins forming each dimer. After Fourier transform, the structure factor can be written as 
\begin{equation}
    S^{\alpha \beta}(\mathbf{q})  = \sum_{i',j'} \sum_{a,b}e^{i \mathbf{q}(\mathbf{r}_{j'b} - \mathbf{r}_{i'a})} \langle S^{\alpha}_{i'a}(\mathbf{q}) S^{\beta}_{j'b}(-\mathbf{q}) \rangle 
\end{equation}
where $\mathbf{r}_{i'a}$ denotes the position of the spin with respect to the unit cell center. In further deriving this expression, the structure factor separates into a quadratic contribution of the form $\bar{s}^2 \tilde{t}_{i}^{\dagger}\tilde{t}_{j}$, and a quartic contribution of the form $\tilde{t}_{i}^{\dagger}\tilde{t}_{j}\tilde{t}_{k}^{\dagger}\tilde{t}_{l}$. In this paper quartic interactions have been neglected, as their contribution is found to be small and does not qualitatively affect the results discussed below. Moreover, the structure factor gives features within the enlarged VBS unit cell, so it has a larger periodicity than the first BZ. As a result, the full momentum-space structure of $S(\mathbf{q})$ is only revealed when plotted over multiple BZs. In practice, this is achieved by shifting the momentum $\mathbf{q}$ by reciprocal lattice vectors to assemble an extended BZ representation.

We compute the trace of the static structure factor $S^{xx}(\mathbf{q})+S^{yy}(\mathbf{q})+S^{zz}(\mathbf{q})$, which we denote $S(\mathbf{q})$, for $d_z = 0.18$ and zero magnetic field. However, we find that the static structure factor is unchanged compared to the case without DMI ($d_z = 0$). This indicates that, within the present mean-field treatment, DMI does not have an apparent effect on the mean field ground state equal-time spin correlations, which remain dominated by the underlying singlet background of the VBS state.
Fig.~\ref{SSF}(b) shows a total of 64 BZs, while the two possible chiral pinwheel configurations present in the crystal have been averaged over. Their separate contributions are shown in Fig.~\ref{SSF}(a). Notably, Fig.~\ref{SSF}(a) shows that the static structure factor reflects the underlying crystal chirality of the dimers.

The resulting structure factor, shown in Fig.~\ref{SSF}(b), exhibits no sharp Bragg peaks. Instead, the intensity is distributed into “triplet” clusters of maxima forming a three-peak structure arranged with sixfold rotational symmetry. This is surprising, as one might expect a dominant peak at the $\Gamma$ point, or its symmetry equivalents, as this is where the triplon branch condenses to and the three-peak structure persists up to fields close to the triplon condensation transition. But this result is still consistent with the absence of long-range magnetic order and the presence of a magnetically disordered ground state.

Furthermore, a strong minimum is observed at the $\Gamma$ point. This suppression at $\mathbf{q} = \Gamma$ can be understood from the dominant intra-dimer correlations. Indeed, for a system of uncorrelated singlet dimers, the on-site contribution is $\langle \mathbf{S}_i \cdot \mathbf{S}_i \rangle = 3/4$ and the intra-dimer contribution is $\langle \mathbf{S}_L \cdot \mathbf{S}_R \rangle = -3/4$, leading to a simplified structure factor of the form $S(\vec{q}) \propto \frac{3}{4}\cdot(1- \frac{1}{2}\sum e^{i\vec{q}(\vec{r_j}-\vec{r_i})}) \propto 1-\cos(\vec{q}\cdot \vec{r})$, which vanishes quadratically for small $\mathbf{q}$. The three peak structure arises once inter-dimer interactions are included. 
\begin{figure*} [t]
\includegraphics[width=0.99\textwidth]{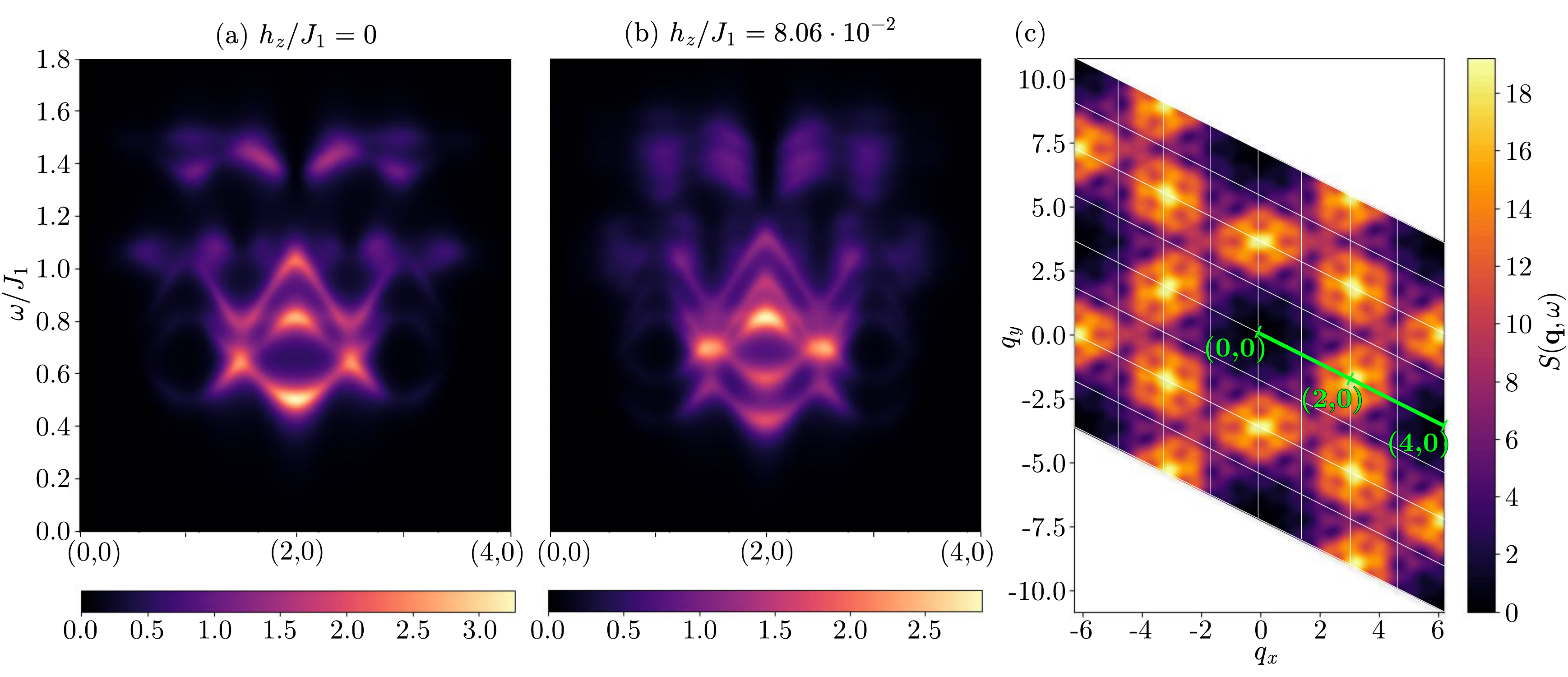}
\caption[width=1\textwidth]{\em Dynamical structure factor $S(\mathbf{q},\omega) = S^{xx}(\textbf{q}, \omega) + S^{yy}(\textbf{q}, \omega) + S^{zz}(\textbf{q}, \omega)$ at zero temperature. (a) Intensity along a one-dimensional momentum path (indicated in green in panel (c)) for $d_z = 0.18$ and zero magnetic field. (b) Same as in (a) in the presence of an external magnetic field of intensity $h = 1.5$ meV applied along the $z$ direction, showing the Zeeman splitting of the triplon branches. (c) dynamical structure factor intensity integrated over the low-energy window $0.4 \leq \omega \leq 1.0$ with no magnetic field, plotted in an extended Brillouin zone scheme obtained by assembling 64 zones. The results are averaged over the two pinwheel domains.}
\label{DSF plots}
\end{figure*}

\subsection{Dynamical Structure Factor}
The dynamical structure factor is defined as
\begin{equation}
    S^{\alpha \beta}(\textbf{q}, \omega) = \frac{1}{N_s} \sum_{m,n}e^{i \textbf{q}\cdot(\textbf{r}_m - \textbf{r}_n)} \int^{\infty}_{-\infty} dt e^{-i \omega t} \langle {S}^{\alpha }_m(t){S}^{\beta}_n(0) \rangle.
\end{equation}
As for the static structure factor, within the bond-operator mean field framework, one can substitute the triplet operators and as a results, $S^{\alpha \beta}(\textbf{q}, \omega) $ naturally separates into a quadratic contribution and a quartic contribution. The dynamical structure factor is computed at zero temperature and refers to the trace of $S^{\alpha \beta}(\textbf{q}, \omega)$, which we denote $ S(\textbf{q}, \omega) = S^{xx}(\textbf{q}, \omega) + S^{yy}(\textbf{q}, \omega) + S^{zz}(\textbf{q}, \omega) $. The full derivation is provided in Appendix~\ref{app_DSF}. Furthermore, all the results presented here are obtained at the quadratic level. In this approximation, the dynamical structure factor only has contributions from single-triplon excitations and is therefore nonzero only for frequencies above the spin gap $\Delta$, i.e. $\omega \geq \Delta$. At zero temperature, higher-order contributions arising from quartic interactions correspond to two-triplon processes and can only start contributing at energies $\omega \gtrsim 2\Delta$. As a result, we find these higher-order processes are strongly suppressed in the low-energy regime and are therefore neglected in this paper.

The dynamical structure factor is averaged over the two possible pinwheel configurations, as both domains are mixed in the crystal. We first consider the dynamical structure factor at fixed frequencies, for the parameters $d_z = 0.18$ and zero magnetic field. Fig.~\ref{DSF plots}(c) shows the sum over a frequency window ranging from $\omega / J_1 = 0.4$ to $\omega/J_1 = 1.0$, corresponding to the low-energy regime. Fig.\ref{DSF plots}(c) was  obtained by assembling 64 BZs and the resulting intensity pattern is in excellent agreement with the numerical result reported in Fig.2(e) of Ref.~\cite{matan2010pinwheel}, which authors calculated via linked-cluster expansion, as a series in interdimer couplings. In particular, the dominant intensity peaks and local minima are located at the same positions in momentum space and repeat with the same periodicity of four times the reciprocal lattice vector.

We also compute the dynamical structure factor along a one-dimensional momentum path, shown in Fig.~\ref{DSF plots}(a). The chosen path (highlighted in Fig.~\ref{DSF plots}(c)) follows that used in Ref.~\cite{matan2010pinwheel} Fig.2(a). The dominant peaks associated with the first and second bands and located at the midpoint of the momentum path [corresponding to the repeated BZ point $(2,0)$ in Fig.~\ref{DSF plots}(c)] are in good agreement with the experimental results. In addition, the overall dispersion and shape of the two lowest-energy bands are well reproduced. In contrast, features associated with higher-energy triplon bands, in particular for frequencies $\omega/J_1 \gtrsim 0.8$, are not displayed in Ref.~\cite{matan2010pinwheel}. Specifically, the third band predicted in our calculation does not appear in the experimental data.

Fig.~\ref{DSF plots}(b) shows the effect of an external magnetic field applied along the $z$ direction. The field lifts the degeneracy between the transverse triplon modes, leading to a clear splitting of the spectral branches. This splitting is directly reflected in the dynamical structure factor as a separation of intensity peaks, consistent with the Zeeman splitting of the underlying triplon bands.

Lastly, for completeness, we also computed the dynamical structure factor along an additional momentum path used in Ref.\cite{matan2010pinwheel} and the result is presented in Appendix \ref{app_DSF}.

\section{\label{sec:THE}Band topology and thermal Hall current}
The topological properties of the triplon bands are characterized through the computation of Berry curvatures and associated band Chern numbers. These quantities are evaluated numerically using the computational discretized plaquette (link-variable) method introduced in Ref.~\cite{fukui2005chern}. This construction ensures gauge invariance and numerical stability of the resulting topological invariants, provided that the bands are non-degenerate, i.e. $|E_n(\mathbf{k}) - E_{n\pm1}(\mathbf{k})| \neq 0$. A detailed description of the computation and evaluation of Berry curvature for bosonic quasiparticles within the BdG formalism is provided in Appendix C of Ref.~\cite{d2024kitaev}. 

\subsection{Band Berry curvatures and Chern numbers}
The main issue encountered in the study of the bands' topology arised from the fact that, while the magnetic field lifts the degeneracy between the three triplet channels, bands originating from different triplon trios may intersect. At such intersections, the Berry curvature, and therefore the band Chern number, are ill-defined. To overcome this issue we introduced a small momentum-independent perturbation to lift the accidental degeneracies and the details are given in Appendix \ref{app_topology}.

We now turn to the results for the band topology. When the bands are well separated, they can carry nontrivial Chern numbers. In particular, the two lowest energy bands exhibit Chern numbers equal to $\pm 1$, as highlighted in Fig.\ref{energy bands} (c). The corresponding Berry curvatures are strongly concentrated near the X-point of the BZ. Moreover, as the magnetic field is varied, multiple bands cross and this leads to various topological phase transitions characterized by different sets of Chern numbers. Notably, these numerous exchanges seem to allow for relatively large Chern numbers. While high Chern numbers have also been reported in related triplon systems~\cite{d2024kitaev, esaki2505spin}, in the present case their precise values may partially depend on the numerical perturbation introduced to lift accidental degeneracies and therefore needs to be interpreted with caution.
We also evaluated Chern numbers for groups of bands using the generalization of the link-variable method presented in Ref.~\cite{fukui2005chern}. The results are trivial, meaning nontrivial topology only emerges upon separating the individual triplon bands, i.e. after lifting the degeneracy between the three triplet channels.

Triplons can be viewed as mobile excitations propagating on the singlet dimer background. More precisely, they correspond to a linear combination of triplet excitations. In the present system, the dimers form an effective lattice, formed by connecting the center of all nearest neighbor dimers and often referred to as a ruby lattice, on which the triplons hop. In the absence of DMI and magnetic field, the Hamiltonian is real-valued, meaning all hopping amplitudes are real and no Berry phase can be accumulated along closed paths. As noted in Sec.~\ref{Energy bands}, the magnetic field splits the triplon bands uniformly by adding a constant on-site term $\pm h \sum_i t_{\pm,i}^\dagger t_{\pm,i}, \ t_\pm \sim t_x \pm it_y $. This removes the spin degeneracy and the eigenstates are given in terms of $t_\pm$ and $t_z$. However despite breaking TRS, it has no effects on the triplon hopping processes and therefore the Berry curvature. In contrast, the DMI generate complex hopping amplitudes and allow the quasiparticles to acquire finite Berry phases when going around closed paths in the lattice.

\begin{figure} []
\centering
\includegraphics[width=1\linewidth]{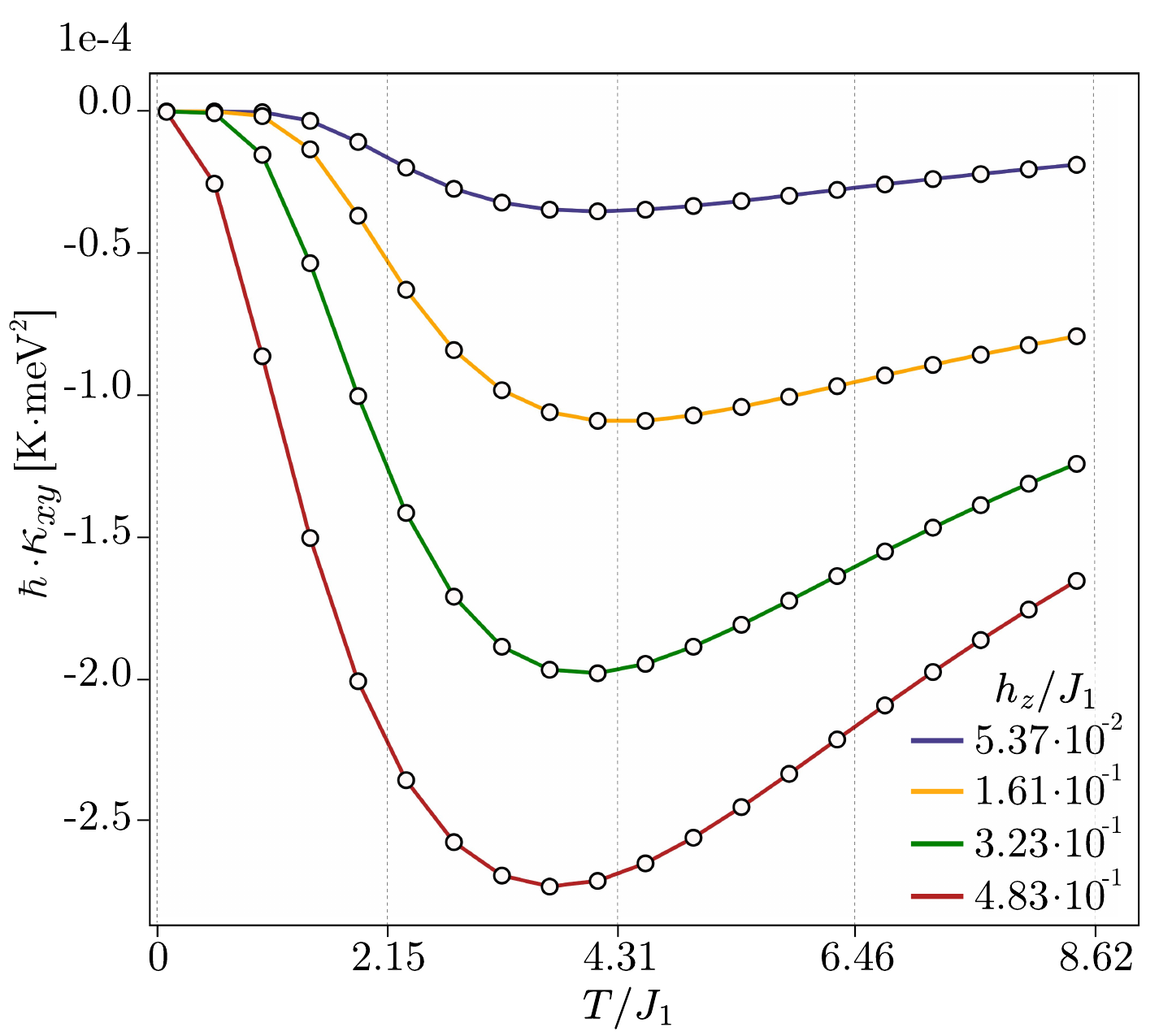}
  \caption{\em Transverse thermal Hall conductivity $\kappa_{xy}$ as a function of temperature for several values of applied magnetic field in the z-direction.}
\label{THE}
\end{figure}
\subsection{Thermal Hall current}
Because the triplons bands have a finite Berry curvature, a temperature gradient leads to a transverse thermal Hall current. The transverse thermal Hall conductivity, $\kappa_{xy}$, is computed using the formula given in Ref.\cite{matsumoto2011theoretical}, which expresses $\kappa_{xy}$ as a Berry-curvature-weighted sum over bosonic bands :
\begin{equation}
    \kappa_{xy} = \frac{k_B^2T}{\hbar L} \sum_{n, \mathbf{q}} c_2[\rho_n(\mathbf{q})]\Omega_{n}(\mathbf{q}),
\label{eq THE}
\end{equation}
with $\rho_n$ the Bose distribution, $\Omega_{n}$ the Berry curvature of the nth band, $c_2(\rho) = (1+\rho)\log^2(\frac{1+\rho}{\rho}) - \log^2(\rho) - 2 \text{Li}_2(-\rho)$, Li$_2$ the polylogarithm function, $T$ the temperature, and $L$ the number of momentum points. This formulation updates earlier results based on the Kubo formula~\cite{katsura2010theory} by including the contribution from the self-rotation of the excitations wave packets, and allows to compute the value of the Thermal Hall transverse coefficient as a function of the temperature.

Fig.\ref{THE} shows $\kappa_{xy}$ as a function of the temperature, for a temperature going from $0$ to $100$ K, and for different values of positive magnetic field in the z-direction. At low temperatures, the triplon population is exponentially suppressed due to the finite spin gap, resulting in a negligible $\kappa_{xy}$. A clear onset of the thermal Hall signal is observed as the temperature increases.  Indeed, as the temperature increases, the lowest-energy bands become thermally populated, leading to a rapid increase in the thermal Hall response.

The magnitude of $\kappa_{xy}$ grows with the applied magnetic field. This can be understood from the fact that the field lowers the energy of the lowest triplon band, thereby increasing its thermal occupation. Since the lowest band has a non-zero Chern number and carries a Berry curvature, its increased population results in a larger contribution to $\kappa_{xy}$. Importantly, the dominant contribution originates from the lowest triplon band, because of the weighting factor $c_2$ in the thermal Hall conductivity formula Eq.\eqref{eq THE}, which favors low-energy excitations.

From our calculation, we observe that $\kappa_{xy}$ changes sign when changing the sign of the magnetic field. This reflects the redistribution of Berry curvature among the triplon bands, as depending on the magnetic field sign, the magnetic field selectively lowers one of the transverse ($x$ or $y$) triplon branches. In particular, the two lowest energy bands are x and y branches with opposite Chern numbers $C = \pm 1$, which leads to a competition between their contributions.

Furthermore, upon increasing temperature, $|\kappa_{xy}|$ reaches a maximum, and subsequently decreases at higher temperatures. Averaging over the field range $h_z = 0.2$ to $8.0$ meV, we obtain an average temperature value of approximately $46.45$ K. This maximum arises from a competition between the increasing thermal occupation of triplon bands and the progressive cancellation of contributions from bands with opposite Berry curvature at higher energies. 
Note that the temperature at which $|\kappa_{xy}|$ is maximal varies only weakly with magnetic field. We attribute this behavior to the fact that the magnetic field shifts the transverse triplon branches approximately uniformly across the spectrum. More specifically, within each trio of bands, one transverse branch is lowered by the field in a similar manner as the lowest-energy band.
As a consequence, although the lowest band moves closer to zero energy, its separation from the other transverse bands with the same spin projection remains relatively unchanged. In addition, as commented in the previous section, the Berry curvature of the lowest band is concentrated near the X point rather than near the condensation point at $\Gamma$, where the band is energetically closer to neighboring transverse branches. This further enhances the role of nearby bands in the thermal Hall response and contributes to the weak field dependence of the temperature at which $|\kappa_{xy}|$ is maximal.

Overall, these results demonstrate that a finite thermal Hall response emerges naturally from the nontrivial topology of the triplon bands, and that its magnitude is tunable with an applied magnetic-field.

\section{Discussion and Conclusion}
In this work, we have investigated the properties of triplon excitations in the pinwheel VBS on the deformed kagome lattice within the bond-operator mean-field theory. Starting from a spin model including Heisenberg exchange, DMI, and an external magnetic field, we derived the triplon BdG Hamiltonian and computed a range of experimentally relevant observables. The validity of the bond-operator description was assessed through the calculation of dynamical structure factors and we found good qualitative agreement of the features of the low-energy spectrum with neutron scattering results. The static structure factor was also computed and it displays a three-peak intensity pattern.

Building on this agreement with experiment, we investigated the topological properties of the triplon bands. At the level of the excitation spectrum, we showed that the chirality of the pinwheel VBS structure alone does not generate nontrivial topology. Indeed, while a magnetic field lifts the degeneracy between triplon branches, it cannot open topological gaps. In contrast, DMI play a crucial role by effectively generating complex hopping amplitudes for the triplons, thereby acting as a source of Berry curvature. We computed the Chern numbers of the individual bands and found that isolated bands can carry nontrivial Chern numbers, with the lowest bands exhibiting values of $\pm 1$, and higher bands reaching larger values through successive band crossings and Berry curvature redistribution. Importantly, the lowest band can be isolated from the rest of the spectra via an applied magnetic field. We then connected the topological properties to the thermal Hall conductivity. Our results show a finite transverse thermal response arising from the Berry curvature of the triplon bands. The thermal Hall signal exhibits a clear onset with increasing temperature and its magnitude increases with magnetic field.

It is important to emphasize that our results are qualitative rather than quantitative. The bond-operator mean-field theory significantly overestimates the spin gap, which we expect to in turn shift the thermal activation temperature to higher temperatures. As a consequence, the obtained average temperature of $46.45$ K, at which the thermal Hall signal is maximal, should be lower in actual materials. A similar comment can be made of the magnetic field applied here. In our computation, the maximal field used would greatly exceed the condensation field observed experimentally~\cite{matan2010pinwheel}.

Another important observation is that it is experimentally observed that there is thermal damping of the spin excitation, leading to a suppression of well-defined excitation peaks for a temperature $30$ K~\cite{matan2010pinwheel}. Since the thermal Hall effect relies on coherent quasiparticle transport, such damping is expected to reduce the observable signal at high temperatures. Furthermore, note that temperature enters our calculation only through the thermal Hall conductivity formula via the Bose occupation factors. The mean-field parameters and triplon spectrum are obtained from a zero-temperature calculation and are assumed to remain unchanged with temperature. In reality, thermal fluctuations would renormalize both these parameters, potentially modifying the quantitative temperature dependence of the thermal Hall response. For all these reasons we expect that a clear thermal Hall signal appears well below ~40 K and below the condensation field of 21 T \cite{matan2010pinwheel}.

As a future research direction, it could be interesting to go beyond the mean-field treatment as employed here. While it successfully captures the qualitative features of the band structure and topology, it does not account for interaction-driven effects that can play a decisive role in transport properties of triplons~\cite{suetsugu2022intrinsic, cairns2020thermal,koyama2026impact} and topological magnon systems~\cite{habel2024breakdown,koyama2024thermal}. In particular, incorporating triplon–triplon interactions explicitly, as done in other analytical approaches to triplon systems developed in Refs.~\cite{kotov1998novel, sushkov1998bound,koyama2026impact}, could be a natural direction for future work. Such extensions would provide a more accurate description of both spectral and transport properties, and help further clarify the conditions under which thermal Hall current signatures can be robustly observed experimentally. Finally, understanding the effect of extrinsic contributions, like side jump~\cite{mangeolle2026extrinsic}, for triplon excitations will be important for making quantitative predictions. 

\section*{Code availability}
The numerical code used to generate the results presented in this work is available at DOI: 10.5281/zenodo.20917612.

\section*{Acknowledgments}
LC thanks Jonas Habel and Pratyay Ghosh for helpful discussions. We acknowledge support from the Deutsche Forschungsgemeinschaft (DFG, German Research Foundation) under Germany’s Excellence Strategy (EXC–2111–390814868 and ct.qmat EXC-2147-390858490), and DFG Grants No. KN1254/1-2, KN1254/2-1 TRR 360 - 492547816 and SFB 1143 (project-id 247310070), as well as the Munich Quantum Valley, which is supported by the Bavarian state government with funds from the Hightech Agenda Bayern Plus. JK thanks the Keck foundation for support.
FM acknowledges support from the Swiss National Science Foundation under grant number 212082.

\clearpage

\appendix
\section{\label{sec:triplonMF}Bond-operator formalism and mean-field Hamiltonian}
\label{app:bondop}
\begin{figure} []
\centering
\includegraphics[width=0.9\linewidth]{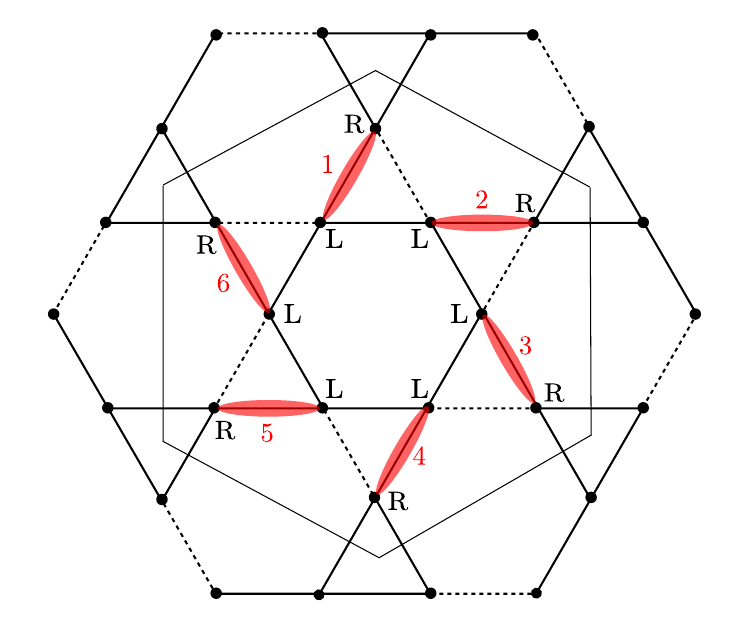}
  \caption{\em Dimerization pattern of the pinwheel VBS. Dimers (in red) are indexed, and each spin is labeled as left (L) or right (R) within the unit cell.}
\label{left right}
\end{figure}

In this Appendix, we summarize the main steps of the bond-operator mean-field formulation used to obtain the triplon spectrum, while referring to Refs.~\cite{sachdev1990bond, hwang2012influence} for further details.
The dimer covering and notation used throughout are shown in Fig.~\ref{left right}. Each dimer is assigned an index, and the two spins forming a singlet are labeled as left (L) and right (R).
\subsection{Bond-operator representation}
Spin operators are expressed in terms of bosonic singlet and triplet operators in the rotated basis (see Eqs.~\eqref{bond_states} and \eqref{bond_states rotated}). In this representation, the spin operator on site $i$ reads
\begin{equation}
\begin{split}
    S_i ^{\alpha} = & \frac{i}{2}(c_{i}(\tilde{s}^{\dagger}{t}_{\alpha}\cos\theta + \tilde{t}_z^{\dagger}{t}_{\alpha} \sin\theta  - h.c.)- \mathcal{E}_{\alpha \beta \gamma}{t}_{\beta}^{\dagger} {t}_{\gamma}),
\end{split}
\end{equation}
with $c_{i} = +1$ ($-1$) for right (left) spins, and the original triplet operators are related to the rotated ones via $t_{x/y}^{\dagger} = \tilde{t}_{x/y}^{\dagger}$ and $t_z^{\dagger} = -\tilde{s}^{\dagger}\sin\theta + \tilde{t}_z^{\dagger}\cos\theta$. 

The full Hamiltonian can be decomposed as
\begin{equation}
    \mathcal{H} = \mathcal{H}_{\text{Dimer}}+ \mathcal{H}_{\mu} + \mathcal{H}_J + \mathcal{H}_D + \mathcal{H}_h .
\end{equation}
Specifically, the local dimer Hamiltonian is diagonal in the rotated basis:
\begin{equation}
    \mathcal{H}_{Dimer } = \sum_{i \in \mathcal{D}} \Big[ E_s \tilde{s}_i\tilde{s}_i^{\dagger} + \sum_{\alpha = x,y,z} E_{\alpha} \tilde{t}_{i \alpha}\tilde{t}_{i \alpha}^{\dagger} \Big]
\end{equation}
with the eigen-energies $E_{\tilde{s}} = -\frac{1}{4}J_1-\frac{1}{2} \sqrt{J_1^2 + d_z^2},$ $E_z = -\frac{1}{4}J_1 + \frac{1}{2} \sqrt{J_1^2 + d_z^2}$, and $E_x = E_y =  \frac{1}{4}J_1$.

The inter-dimer Heisenberg term $\mathcal{H}_J$ and DMI term $\mathcal{H}_D$ generate quadratic and quartic interactions between triplons. Their explicit forms are given below :
\begin{equation}
\begin{split}
&\mathcal{H}_J = \sum_{\langle ij \rangle'} \frac{J_{ij}}{4} \sum_{\alpha = x,y,z} \Bigg[C_{ij,\alpha}^{(1)}(-\tilde{s}_i\tilde{s}_j\tilde{t}^{\dagger}_{i \alpha }\tilde{t}^{\dagger}_{j \alpha } + \tilde{s}^{\dagger}_i\tilde{s}_j\tilde{t}^{\dagger}_{i \alpha}\tilde{t}_{j \alpha } + h.c.)
\\ \\ &
+ \frac{1}{2}\sum_{ \beta \neq \alpha}C_{ij,\alpha \beta}^{(2)}(-\tilde{t}^{\dagger}_{i\alpha }\tilde{t}^{\dagger}_{j\alpha }\tilde{t}_{i\beta }\tilde{t}_{j \beta } + \tilde{t}^{\dagger}_{i \alpha }\tilde{t}_{j \alpha }\tilde{t}_{i \beta }\tilde{t}^{\dagger}_{j \beta } + h.c.) \Bigg],
\label{H_J eq}
\end{split}
\end{equation}
where $\langle ij \rangle'$ denotes the inter-dimer nearest neighbour pairs of spins, $i$ and $j$ being the indices of the singlet to which the spin belongs to, $C_{ij, \alpha}^{(1)} = (c_i c_j \cos^2(\theta) + \sin^2(\theta))$ for $\alpha = x,y$ and $C_{ij, \alpha}^{(1)} = c_i c_j$ for $\alpha = z$, $C_{ij,\alpha \beta}^{(2)} =1$ if the pair $\alpha \beta$ is $x$ and $y$, else $C_{ij,\alpha \beta}^{(2)} =(c_i c_j \sin^2(\theta) + \cos^2(\theta))$, and :
\begin{equation}
\begin{split}
    & \mathcal{H}_D = \sum_{\langle ij \rangle'}
    \frac{D_{ij}}{4}\Bigg[  C_{ij,x}^{(1)}((-\tilde{s}_i \tilde{s}_j\tilde{t}^{\dagger}_{i x}\tilde{t}^{\dagger}_{j y} + \tilde{s}_i \tilde{s}_j^{\dagger}\tilde{t}^{\dagger}_{i x}\tilde{t}_{j y} + h.c.) + \\ & (\tilde{s}_i \tilde{s}_j\tilde{t}^{\dagger}_{i y}\tilde{t}^{\dagger}_{j x} - \tilde{s}_i \tilde{s}_j^{\dagger}\tilde{t}^{\dagger}_{i y}\tilde{t}_{j x} +h.c. ) ) \\ \\ &
    +C_{ij,xz}^{(2)}((- \tilde{t}^{\dagger}_{i z}\tilde{t}^{\dagger}_{j z}\tilde{t}_{i x}\tilde{t}_{j y} + \tilde{t}^{\dagger}_{i z}\tilde{t}_{jz}\tilde{t}_{ix}\tilde{t}^{\dagger}_{j y} +h.c.) \\ & 
    + (\tilde{t}^{\dagger}_{iz}\tilde{t}^{\dagger}_{jz}\tilde{t}_{iy}\tilde{t}_{jx} - \tilde{t}^{\dagger}_{iz}\tilde{t}_{jz}\tilde{t}_{iy}\tilde{t}^{\dagger}_{jx} + h.c.)) \\ \\ &
    +C_{ij}^{(3)} (\tilde{s}_i \tilde{s}_j\tilde{t}^{\dagger}_{i\alpha }\tilde{t}^{\dagger}_{j\alpha } -\tilde{s}_i \tilde{s}_j^{\dagger}\tilde{t}^{\dagger}_{i\alpha }\tilde{t}_{j\alpha } + h.c.) \\ &
    +C_{ij}^{(3)} ( \tilde{t}^{\dagger}_{iz}\tilde{t}^{\dagger}_{jz}\tilde{t}_{i\alpha }\tilde{t}_{j\alpha } - \tilde{t}^{\dagger}_{iz}\tilde{t}_{jz}\tilde{t}_{i\alpha }\tilde{t}^{\dagger}_{j\alpha } + h.c.) \Bigg]
\end{split}
\end{equation}
with $C_{ij}^{(3)} =(c_j - c_i)\cos(\theta) \sin(\theta)$.
\subsection{Mean-field approximation}
Quartic triplon interactions are treated within a Hartree–Fock mean-field approximation. The singlet operator is replaced by its expectation value $\tilde{s} \to \bar{s}$, and quartic terms are decoupled using the mean fields defined in Eq.~\eqref{MF}. The parameters $\mu$ and $\bar{s}$ are determined self-consistently from the saddle-point equations:
\begin{equation}
    \left\langle \frac{\partial \mathcal{H}_{MF}}{\partial \mu} \right\rangle = 0 \text{, } \left\langle \frac{\partial \mathcal{H}_{MF}}{\partial \bar{s}} \right\rangle {=} 0 
    \label{self-consistency eq}
\end{equation}
\subsection{Momentum-space formulation}
Denoting $\mathbf{q}$ the momentum in the first BZ, the Hamiltonian is Fourier transformed using
\begin{align}
\sum_{\mathbf{r}} t_{i \alpha}^{\dagger}(\mathbf{r}) t_{j \beta}(\mathbf{r}+\mathbf{R}) &= \sum_{\mathbf{q}} e^{-i \mathbf{q}\cdot \mathbf{R}} t_{i \alpha }^{\dagger}(\mathbf{q}) t_{j \beta }(\mathbf{q}), \\
\sum_{\mathbf{r}} t_{i \alpha }(\mathbf{r}) t_{j\beta}(\mathbf{r}+\mathbf{R}) &= \sum_{\mathbf{q}} e^{i \mathbf{q}\cdot \mathbf{R}} t_{i\alpha}(\mathbf{q}) t_{j \beta }(-\mathbf{q}).
\end{align}
The various contributions to the Hamiltonian reduce to the following forms.
\begin{equation}
\begin{split}
    &\mathcal{H}_{\text{Dimer}} + \mathcal{H}_{\mu} = \frac{N_s}{2}(E_s \bar{s}^2 +\mu (1-\bar{s}^2)) +\\ &  \sum_{\mathbf{q}, \alpha} \frac{E_{\alpha}-\mu}{2}(e^{-i \mathbf{q} \cdot \mathbf{R}}t_{i \alpha}^{\dagger}(\mathbf{q})t_{i \alpha}(\mathbf{q})+e^{i \mathbf{q} \cdot \mathbf{R}}t_{i \alpha}^{\dagger}(-\mathbf{q})t_{j \alpha}(-\mathbf{q}))
\end{split}
\end{equation}
\begin{equation}
\begin{split}
    &\mathcal{H}_{h_z} =\\& -\sum_{\mathbf{q}} i h_{z}\mathcal{E}_{z \alpha \beta} (e^{-i \mathbf{q} \cdot \mathbf{R}}t_{i \alpha}^{\dagger}(\mathbf{q})t_{j \beta}(\mathbf{q}) + e^{i \mathbf{q} \cdot \mathbf{R}}t_{i \alpha}^{\dagger}(-\mathbf{q})t_{j \beta}(-\mathbf{q}))
\end{split}
\end{equation}
\begin{equation}
\begin{split}
&\mathcal{H}_J = \sum_{\mathbf{q}} \frac{J_{ij}}{8} \sum_{\alpha = x,y,z}\Bigg[ C_{ij, \alpha} ^{(1)}\bar{s}^2(\\ &
- e^{-i \mathbf{q} \cdot \mathbf{R}}t^{\dagger}_{i \alpha}(\mathbf{q})t^{\dagger}_{j \alpha}(-\mathbf{q}) - e^{i \mathbf{q} \cdot \mathbf{R}}t^{\dagger}_{i \alpha}(-\mathbf{q})t^{\dagger}_{j \alpha }(\mathbf{q}) + h.c. \\ &
+ e^{-i \mathbf{q} \cdot \mathbf{R}}t_{i \alpha }^{\dagger}(\mathbf{q})t_{j \beta}(\mathbf{q}) + e^{i \mathbf{q} \cdot \mathbf{R}}t_{i \alpha}(-\mathbf{q})t_{j \beta }^{\dagger}(-\mathbf{q})+ h.c.) 
\\ &
+ \frac{1}{2}\sum_{\beta \neq \alpha}C_{ij,\alpha \beta}^{(2)}(\\&
-Q^{\beta \beta}_{ij}(\mathbf{R})(e^{-i \mathbf{q} \cdot \mathbf{R}}t^{\dagger}_{i \alpha}(\mathbf{q})t^{\dagger}_{j \alpha}(-\mathbf{q})+e^{i \mathbf{q} \cdot \mathbf{R}}t^{\dagger}_{i\alpha }(-\mathbf{q})t^{\dagger}_{j \alpha }(\mathbf{q})) + h.c. \\&
+ P^{\beta \beta}_{ij}(\mathbf{R})(e^{-i \mathbf{q} \cdot \mathbf{R}}t_{i\alpha }(\mathbf{q})t^{\dagger}_{j\alpha }(\mathbf{q}) + e^{i \mathbf{q} \cdot \mathbf{R}}t_{i \alpha}(-\mathbf{q})t^{\dagger}_{j\alpha}(-\mathbf{q}))+ h.c.\\&
+Q^{\alpha \alpha}_{ij} (Q^{\beta \beta}_{ij})^{*}-P^{\alpha \alpha}_{ij} (P^{\beta \beta}_{ij})^{*} + h.c.)\Bigg]
\end{split}
\end{equation}
and $\mathcal{H}_{D}$ has a similar form to $\mathcal{H}_J$.
The Hamiltonian separates into a constant term $\mathcal{H}_0$, depending on the singlet condensate density and the number of sites, and the quadratic bosonic part $\mathcal{H}_t$. 
\subsection{Bosonic Bogoliubov Diagonalisation}
Due to the presence of anomalous terms $t_{i\alpha }t_{j\beta }$ and $t_{i \alpha }^{\dagger}t_{j \beta }^{\dagger}$, the Hamiltonian is diagonalized using a bosonic Bogoliubov–de Gennes (BdG) formalism. Writing 
\begin{equation}
    \mathcal{H}_t = \sum_{\mathbf{q}} \Psi_{\mathbf{q}}^{\dagger}\mathcal{M}(\mathbf{q})\Psi_{\mathbf{q}} 
\label{pre-BDG eq}
\end{equation}
with the bosonic operators gathered in a Nambu spinor,
\begin{equation}
    \Psi_{\mathbf{q}} = (t_{1x}(\mathbf{q}) t_{1y}(\mathbf{q})...t_{6z}(\mathbf{q}) t^{\dagger}_{1x}(-\mathbf{q})t^{\dagger}_{1y}(-\mathbf{q})...)^{T}
\end{equation}
and where the BdG matrix $\mathcal{M}(\vec{q})$ reads
\begin{equation}
    \mathcal{M}(\mathbf{q}) = \begin{pmatrix}
    \mathcal{A}(\mathbf{q}) & \mathcal{B}(\mathbf{q}) \\
    \mathcal{B}(\mathbf{q})^{\dagger}  & \mathcal{A}(-\mathbf{q})^T \\
\end{pmatrix} .
\end{equation}
$\mathcal{A}(\vec{q})$ gathers the coefficients and phases of terms $t_{i\alpha }^{\dagger}(\mathbf{q})t_{j\beta }(\mathbf{q})$ and $\mathcal{B}(\vec{q})$ corresponds to the coefficient and phase of the terms $t_{i\alpha }^{\dagger}(\mathbf{q})t_{j\beta }^{\dagger}(-\mathbf{q})$.

The diagonalization is performed via a paraunitary transformation
\begin{equation}
\Psi(\mathbf{q}) = R(\mathbf{q}) \Phi(\mathbf{q}),
\end{equation}
with
\begin{equation}
R(\mathbf{q}) =
\begin{pmatrix}
\mathcal{U}(\mathbf{q}) & \mathcal{V}(-\mathbf{q}) \\
\mathcal{V}^{*}(\mathbf{q}) & \mathcal{U}^{*}(-\mathbf{q})
\end{pmatrix}.
\end{equation}
$\mathcal{U}$ and $\mathcal{V}$ are $18 \times 18$ matrices. More specifically, the bosonic operators transform as
\begin{equation}
t_{i}(\mathbf{q}) = \sum_{m=1}^{18}\big[ \mathcal{U}_{im}(\mathbf{q})\alpha_{m}(\mathbf{q}) + \mathcal{V}_{im}(\mathbf{q})\alpha_{m}^{\dagger}(-\mathbf{q})\big]
\label{Bogoliubov operator}
\end{equation}
where $i$ labels both the dimer index and spin channel $\alpha$.

The eigenvalue problem
\begin{equation}
\eta \mathcal{M}(\mathbf{q}) \phi(\mathbf{q}) = E(\mathbf{q}) \phi(\mathbf{q})
\end{equation}
then yields the triplon energies and eigenvectors, where $\eta$ is the bosonic metric.

The Hamiltonian is finally diagonalized as
\begin{equation}
\begin{split}
    \mathcal{H}_t & = \sum_{\mathbf{q}} \Phi^{\dagger}(\mathbf{q})\eta \begin{pmatrix}
E_{+}(\mathbf{q}) & 0 \\
0 & -E_{+}(-\mathbf{q})
\end{pmatrix}\Phi(\mathbf{q}) \\ & = \sum_{\mathbf{q}}2 E_{i+}(\mathbf{q})\left[\alpha_i^{\dagger}(\mathbf{q})\alpha_i(\mathbf{q}) + \frac{1}{2}\right]
\end{split}
\end{equation}
which provides the triplon dispersion relations used in the main text.

\section{Derivation of the Dynamical Structure Factor and additional result}
\label{app_DSF}
In this section, we detail the analytical calculation of the dynamical structure factor. The strategy is the following : we compute the imaginary-time spin susceptibility $\chi_{\alpha\alpha'}(\vec q,i\omega_n)$ using Matsubara formalism and obtain the dynamical structure factor using the fluctuation-dissipation theorem
\begin{equation}
S(\vec q,\omega)=\frac{1}{\pi}\big[1+n_B(\omega)\big]\Im \sum_\alpha \chi_{\alpha\alpha}(\vec q,\omega).
\end{equation}
\subsection{Imaginary-time susceptibility}
The imaginary-time spin susceptibility is defined as
\begin{equation}
\begin{split}
    &\chi_{\alpha\alpha'}(\vec q,\tau)
    = \frac{1}{N}
      \sum_{i,j} e^{i\vec q(\vec R_j-\vec R_i)}
      \left\langle T_\tau S_i^\alpha(\tau)\,
      S_j^{\alpha'}(0)\right\rangle 
      \\ &= \frac{1}{N}\sum_{i,j} e^{i\vec q(\vec r_j-\vec r_i)} \int_0^\beta d\tau \ e^{i\omega_n\tau} \left\langle T_\tau S_i^\alpha(\tau) S_j^{\alpha'}(0)\right\rangle 
\end{split}
\end{equation}
where $T_{\tau}$ is the time-ordering operator and which can be expanded in Matsubara frequencies.
Spins are expressed using bond operators and only quadratic and quartic triplon terms are kept. This naturally separates the susceptibility into a quadratic (one-triplon process) contribution and a quartic (two-triplon process) contribution. In the rest of this section we focus on the one triplon processes, as we found the quartic contribution was negligible compared to the quadratic one. Using the dimer index $i'$ associated with spin $i$ and $\alpha$ for the triplet channel, which we take to be homogeneous now, i.e. $\alpha' = \alpha$, the quadratic contribution is 
\begin{equation}
\begin{split}
    & \chi^{(2)}_{\alpha\alpha}(\vec q,i\omega_n)=- \frac{\bar{s}^2}{48}\sum_{i,j} \int_0^{\beta} d \tau e^{i w_n \tau} e^{i \vec{q}(\vec{r}_j - \vec{r}_i)} \\ & \langle T_{\tau} C_{ij, \alpha}^{(1)} \Big(t_{i' \alpha}(-\vec{q}, \tau) t_{j' \alpha} (\vec{q},0)- t_{i' \alpha}(-\vec{q}, \tau)t^{\dagger}_{j' \alpha}(- \vec{q},0) \\ &-  t^{\dagger}_{i' \alpha}(\vec{q}, \tau)t_{j' \alpha}(\vec{q},0) +  t^{\dagger}_{i' \alpha}(\vec{q}, \tau)t^{\dagger}_{j' \alpha}(-\vec{q},0) \Big) .
\end{split}
\end{equation}
Where $\vec{r}_i$ denotes the spin position w.r.t the center of their corresponding unit cell, $C_{ij, \alpha}^{(1)}$ is the constant defined for Eq.\eqref{H_J eq} and $N_u$ is the number of unit cells.
\subsection{Matsubara Green functions}
The normal and anomalous Green functions are defined as
\begin{align}
    G_{ab}^{\alpha\alpha'}(\vec q,i\nu_n)
 &= \int_0^\beta d\tau\; e^{i\nu_n\tau} G_{ab}^{\alpha \alpha'}(\vec{q}, \tau)
       \\ &=-\int_0^\beta d\tau\; e^{i\nu_n\tau}
     \langle T_\tau\, t_{\alpha a}(\vec q,\tau)\,
     t_{\alpha'b}^\dagger(\vec q,0)\rangle ,\\[4pt]
    F_{ab}^{\alpha\alpha'}(\vec q,i\nu_n)
 & = \int_0^\beta d\tau\; e^{i\nu_n\tau} F_{ab}^{\alpha \alpha'}(\vec{q}, \tau)
      \\ &= -\int_0^\beta d\tau\; e^{i\nu_n\tau}
     \langle T_\tau\, t_{\alpha a}(-\vec q,\tau)\,
     t_{\alpha'b}(\vec q,0)\rangle,
\end{align}
where we then take $\beta \rightarrow \infty$. Using the Bogoliubov transformation given in Eq.~\eqref{Bogoliubov operator}, together with thermal averages $\langle \alpha_m^\dagger\alpha_n\rangle =n_B(E_m)\delta_{mn}$ and $\langle \alpha_m\alpha_n^\dagger\rangle =(1+n_B(E_m))\delta_{mn}$, one obtains the pole decomposition
\begin{align}
G_{ab}^{\alpha\alpha'}(\vec q,i\nu_n)
 &= \sum_m\left[
      \frac{\mathcal U_{(\alpha a),m}\mathcal U^*_{(\alpha'b),m}}
           {i\nu_n - E_m}
    - \frac{\mathcal V_{(\alpha a),m}\mathcal V^*_{(\alpha'b),m}}
           {i\nu_n + E_m}\right], \\[4pt]
F_{ab}^{\alpha\alpha'}(\vec q,i\nu_n)
 &= \sum_m\left[
      \frac{\mathcal U_{(\alpha a),m}\mathcal V_{(\alpha' b),m}}
           {i\nu_n - E_m}
    - \frac{\mathcal V_{(\alpha a),m}\mathcal U_{(\alpha' b),m}}
           {i\nu_n + E_m}\right].
\end{align}
The poles at $i \nu_n \pm E_m$ represent the emission and absorption of a triplon.
\subsection{Quadratic (one-triplon) contribution}
Substituting the above Green functions in the quadratic contribution of the susceptibility yields
\begin{equation}
\begin{split}
 & \chi^{(2)}_{\alpha\alpha}(\vec q,i\omega_n)
   = \\& \frac{\bar s^2}{48}
    \sum_{i,j,m} C_{ij, \alpha}^{(1)} e^{i\vec q(\vec r_j-\vec r_i)}
    \Bigg[
     \frac{A^{\alpha}_{i'j',m}(-\vec q)}
            {i\omega_n - E_m(-\vec q)}
    - \frac{(A^{\alpha}_{i'j',m}(\vec q))^*}
            {i\omega_n + E_m(\vec q)}
    \Bigg],
\end{split}
\end{equation}
with $A^{\alpha}_{i'j',m}(\vec q) = \big(\mathcal U_{i'\alpha,m}-\mathcal V^*_{i'\alpha,m}\big) \big(\mathcal U^*_{j'\alpha,m}-\mathcal V_{j'\alpha,m}\big)$.  \\
After analytic continuation $i\omega_n\to\omega+i0^+$, we obtain the retarded susceptibility. Finally, the one-triplon structure factor is
\begin{equation}
\begin{aligned}
& S^{(2)}_{\alpha\alpha'}(\vec q,\omega) = \frac{\bar s^2}{48 \pi}
   \sum_{i,j,m} C_{ij, \alpha}^{(1)} e^{i\vec q(\vec r_j-\vec r_i)} 
   \\& \Im \Bigg(
      \frac{A^{\alpha}_{i'j',m}(- \vec q)}
            {\omega - E_m(-\vec q)+i 0^+}
    - \frac{(A^{\alpha}_{i'j',m}(\vec q))^*}
            {\omega + E_m(\vec q) +i 0^+}
    \Bigg).
\end{aligned}
\end{equation}
At $T=0$, only the $\delta(\omega-E_m)$ pole contributes. Numerically we use a finite broadening, $\omega+i\gamma$ where $\gamma$ is a positive infinitesimal, and we use $\gamma=0.05$.

\subsection{Additional dynamical structure factor result along one-dimensional path}
\begin{figure} []
\centering
\includegraphics[width=1\linewidth]{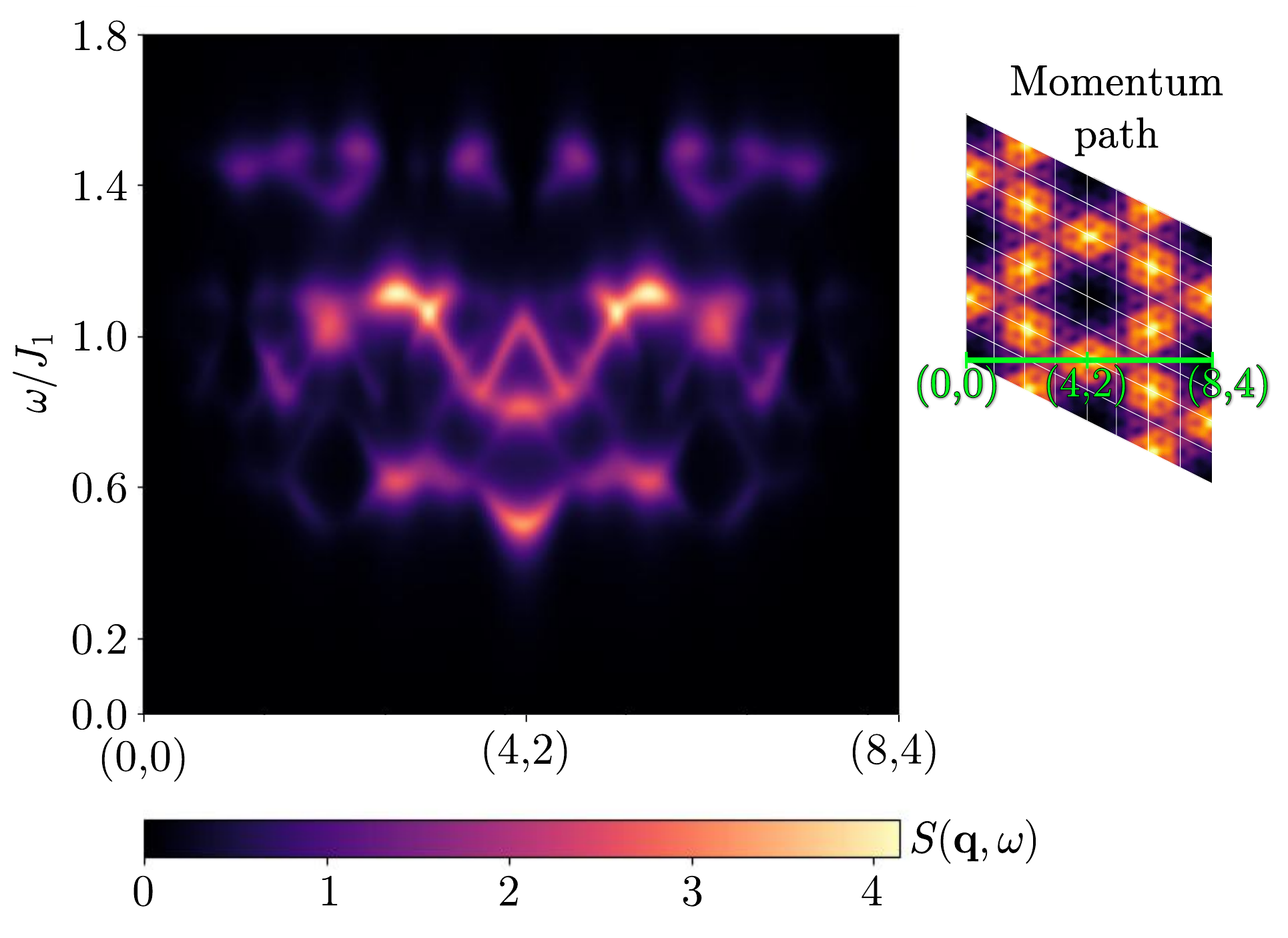}
  \caption{\em Dynamical structure factor $S(\mathbf{q},\omega)$ at zero temperature along a one-dimensional momentum path (indicated in green in the second panel).}
\label{1D DSF path 2}
\end{figure}

In addition to the dynamical structure factor results presented in the main text, we also compute the spectral intensity along an alternative one-dimensional momentum path. The result is shown in Fig.~\ref{1D DSF path 2} for $d_z = 0.18$ and in the absence of a magnetic field. The chosen path, highlighted in green, corresponds to that used in Fig.~2(b) of Ref.~\cite{matan2010pinwheel}.
\\
Consistent with the analysis presented in the main manuscript, the position of the dominant intensity peak of the lowest band, as well as the overall low-energy band structure, are in agreement with the experimental observations. However, features associated with higher-energy triplon bands, are not displayed
in Ref.\cite{matan2010pinwheel}. Specifically, the peaks at an approximate frequency $\omega/J_1$, away from the point (4,2), are not observed in the experimental data \cite{matan2010pinwheel}.

\section{Numerical details of computation of Berry curvatures and Chern numbers}
\label{app_topology}
The BZ was discretized using a $100 \times 100$ momentum grid, which we find sufficient to ensure convergence of the Chern numbers. However, in large regions of parameter space, several triplon bands cross and are therefore not well isolated. In these situations, the Berry curvature is not well defined for individual bands, leading numerically to singularities in the Berry curvature at band-touching points and the overall Chern numbers not summing to zero.

In order to regularize the problem and enable a stable numerical evaluation of the Berry curvature for each band, we introduce a small momentum-independent Hermitian perturbation to the BdG Hamiltonian.
Concretely, a complex constant is added to all the off-diagonal elements of the BdG Hamiltonian and for all momentum points, in a manner that preserves Hermiticity. More specifically, the perturbation is added to the upper triangular part of the $\mathcal{A}(\mathbf{q})$ block (excluding diagonal elements), with the corresponding complex conjugate added to the lower triangular part, while a uniform contribution is added to all elements of the $\mathcal{B}(\mathbf{q})$ block.

As a result, the accidental degeneracies were lifted without modifying the global topology of the bands. This procedure opens gaps of order $10^{-3}$ meV, which are not visible at the scale of the band structure but are sufficient to regularize the Berry curvatures. Indeed, the Berry curvatures become both smooth and stable, and the Chern numbers satisfy the expected sum rule.

\newpage
\bibliography{biblio}

\end{document}